\documentclass[12pt]{iopart}

\usepackage{iopams}  

\usepackage{graphicx,xr}
\usepackage{color}

\newcommand{\nvec}[1]{{\mathbf #1}}
\begin{document}

\title[Reactive  potentials and geometry]
       {       Reactive interatomic potentials and their geometrical features    }

\author{               Ladislav Kocbach and Suhail Lubbad            }

\address{       Dept. of Physics and Technology, University of Bergen, Norway }
\ead{ladislav.kocbach@ift.uib.no   suhail.lubbad@gmail.com}
%
%%%%%%%%%%%%%%%%%%%%%%%%%%%%%%%%%%%%%%%%%%%%%%%%%%%%%%%%%%%%%%%%%%%%%%%%%%%%
%
%
%
\begin{abstract}
We discuss various approaches to modeling the interatomic interactions for molecular dynamics
with special focus on the geometrical structural properties. The type of interactions considered are so called reactive force fields, i.e. interactions without predefined bonds and structures. The discussed cases cover 
the well known Stillinger-Weber, Tersoff-Brenner, EDIP, ReaxFF and ABOP interaction models as well as some additional examples. We discuss also a recently published synthesis of diamond-like structures by 
isotropic pair potential with multiple minima and use this concepts to propose a sort of classification scheme
for interactions with respect to the geometry modeling. In most details we discuss the Tersoff-Brenner 
potentials and also Stillinger-Weber potential, since these models still appear quite popular in
recent research, even though the newer models are more efficient
in most respects, except of simplicity. We also propose simple modifications of 
the basically three-body interactions in order to attempt the simulation of four-body correlation
effects.
The main motive for this study has been to find how the geometrical features are related to
theoretical concepts and whether all possibilities for simplification are exhausted.
We conclude that there are large variations in the methods to design the empirical interactions,
it does not seem possible to conclude that all possible simple approaches have been considered,
since the mentioned recent multiple minima isotropic pair potential method remained undiscovered until
quite recently. Though this particular method does not have a direct practical importance,
we find its recent discovery as an indication that possible simpler alternatives to 
the existing models should be explored. This paper also forms a basis for our work with new
simple interaction models which is presently submitted for publication.

\end{abstract}

\maketitle

%
%
%
%%%%%%%%%%%%%%%%%%%%%%%%%%%%%%%%%%%%%%%%%%%%%%%%%%%%%%%%%%%%%
%
%
%
%
        \section{             Introduction                }
%
%
%
%%%%%%%%%%%%%%%%%%%%%%%%%%%%%%%%%%%%%%%%%%%%%%%%%%%%%%%%%%
%
%
%
%
%The enormous increase in the activities related to 
%nanostructures and their selfassembly also call for various directions of development in 
%computational studies and modeling. The ab initio methods seem to become
%easier applicable to many problems, but the classical MD with model interactions
%still have their place.
%Molecular dynamics has branched into a variety of approaches differing in
%many aspects. One of the essential features is naturally the 
%way in which the interactions are included.

From some recent papers a reader might get the impression that the empirical 
potentials are no longer necessary because the advances in computing will
soon  allow quantum mechanical calculations of atomic interactions
based on e.g. density functional methods for nearly any type
of atomic scale simulations. This work shows, among other things, that
the empirical potentials are still used in many applications and will 
probably remain to be used for studies of some aspects of particle systems
for a long time.
The project reported here started as a simple investigation of how
do the Tersoff-Brenner potentials     %%%   - or models of interatomic interaction - 
(Tersoff, \cite{Tersoff_1986_Si} \cite{Tersoff_1988_C},   \cite{Tersoff_1988_PRB};
Brenner \cite{brenner1_1990} and second generation \cite{brenner_REBO}) 
model the geometrical or stereochemical
features of the modeled aggregates of atoms. During this work we found
details about a number of alternative approaches
and the comparisons between the various features lead us to the presented
analysis. These results should be useful for researchers starting work
on various molecular structures of the type covered by any of 
the mentioned approaches to atom-atom interactions, as well as for 
projects aiming at development of improved or combined systems.

The popularity of Tersoff-Brenner and the other so called bond-order potentials seems to be based on their 
relative simplicity. The formulae for the potentials contain only elementary functions, 
a set of parameters is available for many situations, and the development of simple computer codes is
quite easy. Additionally, a number of computational implementations are freely available. 
One aspect which seems also important is that 
 they appear as two-body interactions, where the influence of the other atoms is
included in the "bond order" representation.
On the other hand, an analysis of the potentials has not been presented very often. A most complete
discussion which we are aware of is in the paper on "second generation bond order potentials"
where Brenner is one of co-authors. This paper is quite long and contains many interesting views.

The purpose of the present paper is to analyze generally these so called reactive potentials, 
or perhaps reactive model interactions, many of them associated with the term bond order  potentials.
We try to compare their features, 
find out how simple they really are, how well they fulfill the promised function and in
particular to prepare ground for possible further work on new alternatives. In spite of their apparent simplicity 
the work with the potentials is quite convoluted, and the possible simplifications do not appear 
without a critical analysis. It seems that most of the workers in this field do not find
time for such an analysis, since the interest mainly lies in the applications.
% the subject is
%       All of the discussed interaction models belong to a small class of
%       so called reactive potentials, it means that all the atomic species
%       can get rearranged in principle in any way, thus modeling any possible
%       chemical reactions and intermediates, in contrast to the more extensively
%       used "force fields" which involve certain permanent bonds and associated 
%       parameters for their description during the simulations.

% The general approach is from the point of physical rather than chemical 
% angle. We recognize that all the mentioned approaches have contributed
% by modeling certain part of the chemical knowledge, incorporated then
% into a certain physical model, either of potentials  or further additional
% modeling elements.

We address also the question whether the 'bond order' approach is really 
flexible enough to accommodate the 'known chemistry'. The authors of the existing approaches
bring many arguments for the positive answer, but 
it is not easy to see if these are all really valid. In the field of simulations a concept of
"transferability" has been established, which mainly describes the same quality, but puts stress on
the performance rather than what one could call usual scientific criteria.

The paper is organized as follows: In the following section we review 
some example applications and 
different requirements put on the potentials in order to provide a certain reference frame.
In section \ref{potentials_all}
%  (\ref{stillinger_weber_sect} to \ref{ReaxFF_section}) 
we introduce
in both historical  order and increased complexity order the various potentials (or interaction models).
We start with a recent isotropic pair interaction which in spite of isotropy
generates the diamond structure. This gives us the possibility to introduce 
what we call RST-SW axis for classification of the all the interactions discussed.
The following part of this section introduces in \ref{stillinger_weber_sect}.
the well known Stillinger-Weber potential \cite{Stillinger_Weber1985}, followed by
Tersoff potential in section \ref{Tersoff_potential} and six other interaction
models.
 Among the models discussed are interactions known as EDIP, 
ReaxFF, as well as interactions based on training of artificial neural networks  in section \ref{neural_networks}  .
In section  \ref{tersoff_manybody} we analyze the 
Tersoff - Brenner potentials and try to bring them into a more general form of 
a many-body potential.  
We also address the questions why the potentials are not additive and why 
the functional form of exponentials is preferred, especially when other forms
were used  in some earlier works. 
We also shortly comment on the concept of PES (potential energy surface, or rather hypersurface) 
which is a starting point of the works outside of the "bond order" potential approach.
In short,  the potential as approximation to PES is contrasted to the
concept of bond order as starting point.
In section \ref{tersoff_geometry}
we investigate where and how the geometrical features are implemented.
We also discuss the functional shapes of the potentials.   

In section \ref{effective-correlation} we 
discuss  how to possibly add the four-body aspects to the existing models, without 
major redefinitions, also with respect to the modifications of the cut-off
treatment in section \ref{tersoff_cutoff}.

%         (comment: pair forces - pairs of atoms
%          three body forces  - pairs of "pairs", or pairs of bonds
%          four body forces   -  pairs of triplets  (or pairs of angles) )
%
%
%
%1.
%General discussion of the work being done,
%computations, what are the aims of the simulations ...
%
%
%   references to be provided:
%                                quantal  calculations
%                                model potentials
%                              Car Parrinello
%                              Monte - Carlo
%
%%%%%%%%%%%%%%%%%%%%%%%%%%%%%%%%%%%%%%%%%%%%%%%%%%%%%%%%%%
%
%
%
%
\section {  The various applications of molecular simulation methods                     
                                                        \label{methods-and-applications}   }
%
%
%
%
%%%%%%%%%%%%%%%%%%%%%%%%%%%%%%%%%%%%%%%%%%%%%%%%%%%%%%%%%%%

%
%
A review of many various approaches and illustrative examples is given in Binder
 {\it et al}, ref. \cite{MolDynSim}. However, this very nice review by far does not cover all the types of 
applications which one can find in the exploding research in nanoscale sciences. 
Also many monographs exist, we refer here as an example to  a very extensive book on 
 Molecular modeling by  Leach, \cite{Leach_Molec_modeling} with more than 700 pages.
In this section
we will shortly discuss some less usual or even perhaps surprising applications.

Applications of molecular dynamics range from first principles high quality 
quantal  calculations to very simple model potentials  of Lennard-Jones or Morse type
described in detail in the above mentioned reviews. 
There are also differences in the treatment of the mechanics itself (e.g. Car-Parrinello \cite{Car_Parrinello}), but 
main focus is on the classical Newton equations derived using various forms of potentials to
yield the forces acting on the nuclei. For completeness, one should perhaps
mention the Monte-Carlo approaches (review e.g. in  \cite{MolDynSim} or \cite{Leach_Molec_modeling}), 
where the focus is not on the time
development, but on - put simply - walking randomly through geometrical configurations and looking
for the minima in the potential energy.

The Tersoff-Brenner potentials are mainly applied to silicon systems and to carbon compounds, 
including hydrocarbons. The two atoms, C and Si belonging both to the group IV and having a very
similar structure from the point of atomic physics, have very different chemical properties.
This can mainly be associated with different ability to form $\pi$ bonds, which again 
from the point of simple molecular  physics can be illustrated by the two oxides, $SiO_2$ and $CO_2$.
It is thus to some degree surprising that the Tersoff-Brenner potentials can be used for both
silicon and carbon.

In the so called 'Car Parrinello method' (CPMD) \cite{Car_Parrinello} 
one can be trying to replace the model potentials by
quantum chemical results for the electronic energies from density functional theory, usually based
on pseudopotentials for the electronic motion. In a very informative tutorial review of molecular 
dynamics methods \cite{MolDynSim} the authors remark about the CPMD that 
the huge advantage is that one is not relying
on 'often ad hoc' effective interatomic potentials which lack 'any firm quantum chemical foundation'.
This characteristics of the MD-potentials seems appropriate, but it is surprising that the authors
use the wording 'quantum chemical foundation', where simply 'chemical foundation' would be
definitely more appropriate. It is indeed a fact that the quantum chemical methods are 'more fundamental'
than any effective potentials can be, but the ultimate benchmarks are the results of real 
world chemistry and physics, using whichever are the most appropriate experimental and measurement
techniques the various disciplines might provide. 
%

%     In fact, the Tersoff-Brenner type potentials have a history of trying to do exactly this, 
%     adjusting the parameters of the potentials by fitting the known properties of the crystal structures. 
%      The question remains whether the chosen properties are selective enough, since only the
%      parameters and not the forms of the interactions are varied.

%%%%                   
%%%%
%%%%        (Stillinger and Weber 1988 PRL have used 260 atoms )
%%%%                   
%%%%
%%%%

There are many different types of use for the empirical reactive potentials. Some of  them are related to the 
studies of structures of silicon. Silicon has an enormously complicated 
variety of phases in the condensed state, both crystalline and amorphous. The Stillinger-Weber
potential \cite{Stillinger_Weber1985} discussed below has been primarily designed for the study of diamond-like Si structure,
literature on further studies counts possibly hundreds of papers. It has generally been concluded that the 
Stillinger-Weber approach is too simple or rather rigid to model the many possible aggregates of Si atoms.
Another area are the studies
of carbon related structures. One of the most exciting areas are the studies of transition from
graphite to diamond, creation of various types of thin films of carbon structures and other carbon
nanostructures. Here the literature goes possibly to thousands of various studies.
A recent review of the research on carbon nanostructures for advanced composites,
\cite{brenner_progress_phys} reviews also the molecular modeling, with 
many examples of the use of empirical potentials.
%  \cite{brenner_progress_phys} 
%   Carbon nanostructures for advanced composites 
%    Yanhong Hu, Olga A Shenderova, Zushou Hu, Clifford W Padgett and Donald W Brenner 

If classified by methods used, there are investigations based on classical deterministic MD,
papers using Monte-Carlo methods, comparisons of empirical potentials with the 
{\it ab initio} calculations, studies involving processes like irradiation by X-rays 
resulting in rearrangement of the atoms (modification of bonds), bombardment of surfaces.
This short account should be illustrative enough for the observation that there are many
aspects which are sought to be understood and that not all of the current uses can 
be served by the {\it ab initio} methods. The empirical interaction models
will have their use even with increased computational capacity of future hardware.
One example is a very recent (2008) study of transformation of graphite to diamond  
under shock compression \cite{shock_graphite_diamond}, where the main method is the 
Car-Parrinello approach \cite{Car_Parrinello}, but 
Tersoff potential studies are conducted to investigate the role of finite size.
Thus, even if some aspects are examined in {\it ab initio} framework,
the  other aspects might be more straightforward
to simulate in the framework of empirical potentials.

One question which could be addressed also by the empirical potentials is
the existence of the so called cubic and hexagonal versions of the diamond structure.
For diamond, these two phases are well documented and known as 
simply diamond or cubic diamond on one hand, and lonsdaleite, 
or hexagonal diamond on the other hand. This question has been addressed
by DFT studies \cite{First-principlesDiamonds}, but to our knowledge not implemented in the framework of
empirical potentials. 
In sections \ref{tersoff_geometry} and \ref{effective-correlation} we discuss how this can be done.
%In section \ref{tersoffbrenner_papers}
%we discuss, Tersoff's original papers; discussion of Stillinger and Weber;
%Geometric factors - three body potential;
%Tersoff - need for at least four body (we know why) (Lonsdaleite);

%Ultrafast transformation of raphite to diamond: An ab initio study 
%of graphite under shock compression 
%Christopher J. Mundy, Alessandro Curioni,2 Nir Goldman,3 I.-F. Will Kuo,3 
%Evan J. Reed,3 Laurence E. Fried,3 and Marcella Ianuzzi4 
%THE JOURNAL OF CHEMICAL PHYSICS 128, 184701 2008

%
%%%%%%%%%%%%%%%%%%%%%%%%%%%%%%%%%%%%%%%%%%%%%%%%%%%%%%%%%%%%%
%
%
%
%
\section {      Potentials                 \label{potentials_all}   }
%
%
%
%
%%%%%%%%%%%%%%%%%%%%%%%%%%%%%%%%%%%%%%%%%%%%%%%%%%%%%%%%%%%%%
%
In this section we review the most usual reactive potentials, or one should perhaps
use the expression "interatomic interaction models" to cover all of the models.
Simple expression "potentials"
is really appropriate only for the earliest models. This review tries to cover 
most of the different types of models, but it simply can not be
complete since there are many variations in the vast existing literature.
One aspect to look for
is to which degree the method attempts to obtain
the empirical potential as an approximation to PES (potential energy surface). 
This aim is explicitly stated in some formulations. In other approaches,
 this aim remains perhaps implicitly present but explicitly 
other modeling aims are 
expressed. This refers in particular to the explicit reference to "bond order"
in most of the newer approaches discussed here.

We start the discussion by referring to a recent work of
Rechtsman, Stillinger and Torquato  \cite{diamond_by_pair_isotropic}
discussing what they called
"synthetic diamond and wurtzite structures" which are self-assembled 
using only isotropic pair potentials. We shall refer to this work as RST.

The two model potentials given in the paper appear as somewhat weighted
negative functions of the radial distribution function (RDF) for the two
structures. It means there is a general background potential
with narrow minima at the positions where the RDF has peaks.
Clearly, the first local minimum corresponding to the nearest neighbor
must be rather shallow and above all the other minima, otherwise only
a closed packed structure will be the stable one.
%%%%%%%%%%%%%%%%%%%%%%%%%%%%%%%%%%%%%%%%%%%%%%%%%%%%%%%%%%%%%
%
%                         r-----------------n
%                         I  / \    / \     I    FIGURE  
%                         I  | |    | |     I
%                         ^-----------------^
%
\begin{figure}[htb]
\center{
\includegraphics[height=6.5cm]{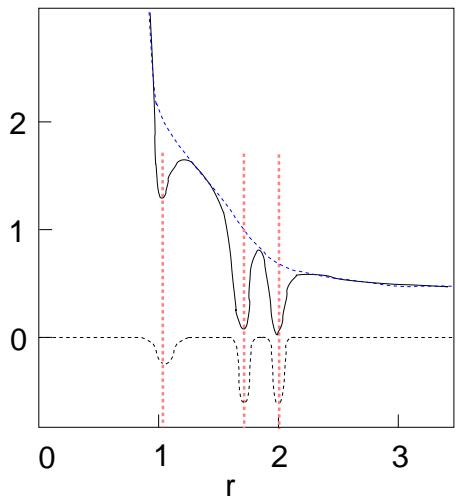}
}
 \caption{Schematic reproduction of the synthetic diamond lattice potential. 
 Full line traces the shape of the potential, curved dashed line the background potential,
 and the lower dotted line the scaled negative of smeared RDF, which is superimposed 
 on the background potential. Note, this is only schematic. It is not a description of
 how the authors 
 constructed the potential.
    }
 %%%
\label{qualit_iso_pot}
\end{figure}
%
%
%
%%%%%%%%%%%%%%%%%%%%%%%%%%%%%%%%%%%%%%%%%%%%%%%%%%%%%%%%%%%%%
Unfortunately, the authors do not describe how they have
arrived to the particular parameterizations, i.e. the extra weights,
but they describe in sufficient detail how they obtain
the positions of the minima, i.e. the distances between the 
next neighbor points of the lattices. 

These potentials are important for our discussion, because here no angular dependence of the forces
between atoms is included, and the potentials are only two-body potentials. This means that 
this interaction model does not carry any explicit geometrical features. All geometry 
is provided by the structure of the Eucledian space itself. This will give us a
possibility to classify the other interactions discussed according to the degree 
of explicit geometrical features. 

When reviewing the potentials, we meet again and again the concept of {\it bond order}, 
and thus we should have a more or less precise definition of this concept. 
That is not easy, since these words are used with at least two different meanings.
IUPAC definition \cite{IUPAC} of bond order can be summarized as follows: considering the region
- it is measure of electron population in the region between atoms A and B which is
moved from the atomic regions of the two centers. In Mulliken's formulation this
is expressed with the help of the electron density matrix. In valence bond theory
the bond order is related to the formal bond orders obtained from the Lewis structure.
The Mulliken's formulation  is related in some way to the electron density, and will be
changing when the configuration is changed. The valence bond 
alternative is related to stable configurations 
and does not have any clear relation to the distance between the two
atomic centers. The exponential relation between the bond strength and 
bond length has been studied by Pauling, also there for stable configurations. 
The exponential form of this dependence has been suggested and in many studies the simple rule has
been confirmed as approximately valid. The relation of bond strength and bond order
are rather unclear, as this reference shows. 

%%%%%% http://goldbook.iupac.org/BT07005.html
%
%%%%%%%%%%%%%%%%%%%%%%%%%%%%%%%%%%%%%%%%%%%%%%%%%%%%%%%%%%%%%
%
%
%
%
\subsection {         Stillinger Weber potential                       \label{stillinger_weber_sect}   }
%
%
%
%
%%%%%%%%%%%%%%%%%%%%%%%%%%%%%%%%%%%%%%%%%%%%%%%%%%%%%%%%%%%%%
%
%
The Stillinger and Weber potential \cite{Stillinger_Weber1985} is the simplest model of non-isotropic interatomic attraction, 
including three body potentials.
It is realized with the help of an isotropic (only distance dependent) two body terms with the addition of 
three-body terms which depend on the angles between two bonds in each of the triplets.
This interaction model includes thus the correlation between any three neighboring atoms
and assures that the preferred bond angle is the so called tetrahedral angle $\tau$
given by $\cos \tau = -1/3$. 
No further considerations of the geometrical relations to other atoms are necessary, 
it is enough to
require the bond angle to be close to 110 degrees. This requirement implicitly defines
the 3 dimensional geometry, since in a plane we can not attach two more atoms to a central atom in an existing triplet and keep all angles close to 110 degrees.
Thus only four neighboring atoms (three bonded to a fourth in the center) can be in one plane with angles 120 degrees, the fifth atom 
(or the fourth closest neighbor of the central atom) must be out of plane, which results in
the rearrangment of the five neighbors into the tetrahedral structure.

The situation is thus similar to the one discussed in the synthetic pair potential case, where however
absolutely no geometry was included explicitly. In Stillinger-Weber case 
only one element of the geometry is included, a preferred bond angle, all the rest is left to the
properties of the Eucledian space. We will classify the remaining potentials by their position
along the RST-SW axis, Where RST is zero and SW is one. 

Stillinger and Weber defined their potential as
%%%%   equation labels 
%%    eq. (\ref{stillinger_TOT_PHI})
%%    eq. (\ref{stillinger_scales})
%%    eq. (\ref{stillinger_f_2})
%%    eq. (\ref{stillinger_f_2})
%%    eq. (\ref{stillinger_h_func})
%
\begin{equation}
\Phi(\nvec{r}_1,\nvec{r}_2,\nvec{r}_3,....,\nvec{r}_N)
=
\sum_{i<j} v_2(\nvec{r}_i,\nvec{r}_j)
+
\sum_{i<j<k} v_3(\nvec{r}_i,\nvec{r}_j,\nvec{r}_k) 
+ ......
\label{stillinger_TOT_PHI}
\end{equation}
and introduced energy and length units, $\varepsilon $ and $\sigma$,
\begin{eqnarray}
v_2(\nvec{r}_i,\nvec{r}_j) 
      &=&       v_2({r}_{ij})
       =        \varepsilon       f_2(r_{ij}/\sigma)
                                                \nonumber \\
v_3(\nvec{r}_i,\nvec{r}_j,\nvec{r}_k) 
      &=&       \varepsilon 
             f_3(\nvec{r}_i/\sigma,\nvec{r}_j/\sigma,\nvec{r}_k/\sigma)
\label{stillinger_scales}
\end{eqnarray}

\begin{equation}
f_2(r)=
    \left\{            %  BIG BRACE
       \begin{array}{l}
        A(Br^{-p}-r^{-q}) \exp\left[  (r-a)^{-1}\right]     , \ \ r<a \\
        0  , \ \ r\ge a 
        \end{array}
    \right. 
\label{stillinger_f_2}
\end{equation}

\begin{equation}
f_3(\nvec{r}_i, \nvec{r}_j, \nvec{r}_k,)=
       h(r_{ij},r_{ik},\theta_{jik})
     + h(r_{ji},r_{jk},\theta_{ijk})
     + h(r_{ki},r_{kj},\theta_{ikj})
\label{stillinger_f_3}
\end{equation}
where $\theta_{ijk}$ is the angle between $\nvec{r}_{ji}$ and $\nvec{r}_{ki}$,
at the vertex $i$. The functions $h$ have two parameters, $(\lambda,\gamma)>0$,
which is nonzero only if both $r_{ij}<a$ and $r_{ik}<a$
\begin{equation}
h(r_{ij},r_{ik},\theta_{jik}) = 
\lambda     \exp\left[   \gamma (r_{ij}-a)^{-1} 
                       + \gamma (r_{ik}-a)^{-1} \right]     
       \ \ 
            \left(  \cos  \theta_{jik}  + \frac{1}{3}     \right)^2
\label{stillinger_h_func}
\end{equation}
and identically equal to zero outside of these two conditions.

They made a limited search, their parameters:
$$
A=7.049556277 , \ \ \ \ B=0.6022245584
$$
$$
p=4, \ \ \ \  q=0, \ \ \ \   a=1.80
$$
$$
\lambda=21.0, \ \ \ \  \gamma=1.20, 
$$
where the units are {\AA}ngstr{\"o}m and electronvolt.
%
%
%    NEEDS      \newcommand{\nvec}[1]{{\mathbf #1}}
%
%%%%%%%%%%%%%%%%%%%%%%%%%%%%%%%%%%%%%%%%%%%%%%%%%%%%%%%%%%%%%
%
%
%                     Using_Stllinger_in_2009_JJAP-48.pdf
%   An Improvement of Stillinger Weber Interatomic Potential Model for 
% reactive ion etching simulations 
% Hiroaki Ohta, Tatsuya Nagaoka, Koji Eriguchi, and Kouichi Ono 
% Japanese Journal of Applied Physics 48 (2009) 020225 RAPID COMMUNICATION 
%
%              \cite{StWe_use_2009} 
%
%%%%%%%%%%%%%%%%%%%%%%%%%%%%%%%%%%%%%%%%%%%%%%%%%%%%%%%%%%%%%
The SW-potentials are mostly considered as only of 
historical interest, However, we can refer to a very recent (2009) study \cite{StWe_use_2009},
where the authors modified and used SW potentials for reactive ion etching simulations.

%%%%%%%%%%%%%%%%%%%%%%%%%%%%%%%%%%%%%%%%%%%%%%%%%%%%%%%%%%%%%
%
%
%
%
\subsection {         Tersoff's potential        \label{Tersoff_potential}      }
%
%
%
%
%%%%%%%%%%%%%%%%%%%%%%%%%%%%%%%%%%%%%%%%%%%%%%%%%%%%%%%%%%%%%
%
%
Tersoff started by defining a two-body interaction, building on the concept of 
bond order, mentioned above. The form of the potential is rather complicated
and it will thus be interesting to classify this interaction 
along the RST-SW axis.
%
%%%%%%%%%%%%%%%%%%%%%%%%%%%%%%%%%%%%%%%%%%%%%%%%%%%%%%%%%%%%%%%%%%%
\begin{equation}\label{eq79} 
E = \frac{1}{2} \sum_{i,j \neq i} V_{ij} ,
\end{equation}
%%%%%%%%%%%%%%%%%%%%%%%%%%%%%%%%%%%%%%%%%%%%%%%%%%%%%%%%%%%%%%%%%%%
%
%
%
%
%            Tersoff
%
%
%
%
%%%%%%%%%%%%%%%%%%%%%%%%%%%%%%%%%%%%%%%%%%%%%%%%%%%%%%%%%%%%%%%%%%%
\begin{equation}\label{eq84}
V_{ij} = f_c(r_{ij}) [ a_{ij} \  A \  \exp (- \lambda_1 \  r_{ij} ) - b_{ij} \  B \  \exp (-\lambda_2 \  r_{ij} )]  
\end{equation}
%%%%%%%%%%%%%%%%%%%%%%%%%%%%%%%%%%%%%%%%%%%%%%%%%%%%%%%%%%%%%%%%%%%
%
%
%
%
%
%
%
%
where $f_c(r)$ is a cut-off function defined below. Here the first term is 
referred to as {\it repulsion}, the second as {\it attraction}. 
The form for $ b_{ij} $ is:
%
%
%
%
%
%
%
%
%%%%%%%%%%%%%%%%%%%%%%%%%%%%%%%%%%%%%%%%%%%%%%%%%%%%%%%%%%%%%%%%%%%
\begin{equation}\label{tersoff-angle}
 b_{ij} =  [1 + ( \beta \zeta_{ij})^n ]^{-\frac{1}{2n}}, \ \ \ \ \ \ \ \ \ \ \ \  \ \ \ \ \ \ \ \ \ \ \ \ \ \ \ \ \ \ \ \ \ \ 
 \end{equation}
%%%%%%%%%%%%%%%%%%%%%%%%%%%%%%%%%%%%%%%%%%%%%%%%%%%%%%%%%%%%%%%%%%%
%
%
%
%
%
%
%
%
%
%
%
%
%
%
%
%
%%%%%%%%%%%%%%%%%%%%%%%%%%%%%%%%%%%%%%%%%%%%%%%%%%%%%%%%%%%%%%%%%%%
\begin{equation}\label{eq86}
\zeta_{ij} = \sum_{k \neq i,j} f_c(r_{ik}) \  g( \theta_{ijk}) \  \exp [{\lambda_3^3 (r_{ij} - r_{ik})}^3],
\end{equation}
%%%%%%%%%%%%%%%%%%%%%%%%%%%%%%%%%%%%%%%%%%%%%%%%%%%%%%%%%%%%%%%%%%%
%
%
%
%
%
%
%
%
%
%%%%%%%%%%%%%%%%%%%%%%%%%%%%%%%%%%%%%%%%%%%%%%%%%%%%%%%%%%%%%%%%%%%
$$ g(\theta) = 1 + \left(      \frac{c}{d}\right       )^2 
                 - 
                  \frac{c^2}{d^2 + [ h - \cos  \theta  ]^2}  
\ \ \ \ \ \ \ \ \ \ \  \ \ \  
$$
%%%%%%%%%%%%%%%%%%%%%%%%%%%%%%%%%%%%%%%%%%%%%%%%%%%%%%%%%%%%%%%%%%%
%
%
%
%
%
%
%
%
%
note that $ b_{ij} \neq  b_{ji} $ - asymmetric formulation.
$ a_{ij} $ proposed form is:
%
%
%%%%%%%%%%%%%%%%%%%%%%%%%%%%%%%%%%%%%%%%%%%%%%%%%%%%%%%%%%%%%%%%%%%
%
%
%
%
%
%
%
%
%%%%%%%%%%%%%%%%%%%%%%%%%%%%%%%%%%%%%%%%%%%%%%%%%%%%%%%%%%%%%%%%%%%%
$$ 
a_{ij} = [1 + ( \alpha \eta_{ij})^n ]^{-\frac{1}{2n}}, \ \ \ \ \ \ \ \ \ \ \ \ \ \ \ \ \ \ \ \ \ 
$$
%%%%%%%%%%%%%%%%%%%%%%%%%%%%%%%%%%%%%%%%%%%%%%%%%%%%%%%%%%%%%%%%%%%%
%
%
%
%
%
%
%
%%%%%%%%%%%%%%%%%%%%%%%%%%%%%%%%%%%%%%%%%%%%%%%%%%%%%%%%%%%%%%%%%%%%  
\begin{equation}\label{eq87}
\eta_{ij} = \sum_{k \neq i,j}   f_c(r_{ik}) \  \exp [{\lambda_3^3 (r_{ij} - r_{ik})}^3]
\end{equation}
%%%%%%%%%%%%%%%%%%%%%%%%%%%%%%%%%%%%%%%%%%%%%%%%%%%%%%%%%%%%%%%%%%%%
%
%
%
If $ \alpha $ is sufficiently small then $ a_{ij} \approx 1 $; Tersoff set $ \alpha $ to zero so that $
a_{ij} = 1 $.
The cut-off function $f_c(r)$ used by all Tersoff followers:
%
%
%%%%%%%%%%%%%%%%%%%%%%%%%%%%%%%%%%%%%%%%%%%%%%%%%%%%%%%%%%%%%%%%%%%% 
\begin{equation}\label{eq85}
f_c(r) =  
\begin{array}{l} 
1 , \ \ \ \ \ \ \ \ \ \ \ \ \ \ \ \ \ \ \ \ \ \ \ \ r < R-D  \\
\frac{1}{2} - \frac{1}{2} \sin[\frac{\pi}{2}\frac{(r - R )}{D}] , \ \ \  R- D < r < R + D \\
0 , \ \ \ \ \ \ \ \ \ \ \ \ \ \ \ \ \ \ \ \ \ \ \ \   r > R + D \\
\end{array} 
\end{equation}
%%%%%%%%%%%%%%%%%%%%%%%%%%%%%%%%%%%%%%%%%%%%%%%%%%%%%%%%%%%%%%%%%%%%
%
%
%

\vskip 1.2cm
The complicated functional dependence of the Tersoff-Brenner potentials 
can be contrasted to that of {\it Stillinger-Weber potential} which is
explicitly of three-body character only, given by the simple
form containing separately two and three body terms
%
%
%
%
%
%
%
%
%%%%%%%%%%%%%%%%%%%%%%%%%%%%%%%%%%%%%%%%%%%%%%%%%%%%%%%%%%%%%%%%%%%
\begin{equation}\label{eq88}
V = 
\frac{1}{2} \sum_{ij} \phi(r_{ij})        + 
\sum_{ijk}
           g(r_{ij})g(r_{ik})
           \left(\cos \theta_{ijk} + \frac{1}{3} \right)^2
\end{equation}
%%%%%%%%%%%%%%%%%%%%%%%%%%%%%%%%%%%%%%%%%%%%%%%%%%%%%%%%%%%%%%%%%%%
%
%
%
%
%
%
%
%
where $ \theta_{ijk} $ is the angle between ${ij} $ and $ {ik} $ bonds, 
$g(r)$ is
a decaying 'cut-off' function.
%%%%%%%%%%%%%%%%%%%%%%%%%%%%%%%%%%%%%%%%%%%%%%%%%%%%%%%%%%%%%%%%%%%
%
%%%%%%%%%%%%%%%%%%%%%%%%%%%%%%%%%%%%%%%%%%%%%%%%%%%%%%%%%%%%%%%%%%%
%
%%%%%%%%%%%%%%%%%%%%%%%%%%%%%%%%%%%%%%%%%%%%%%%%%%%%%%%%%%%%%%%%%%%
%
%
%
%
%
%
%
%%%%%%%%%%%%%%%%%%%%%%%%%%%%%%%%%%%%%%%%%%%%%%%%%%%%%%%%%%%%%
%
%
%
%
\subsection {        REBO - Second Generation Brenner Potentials       \label{REBO_potential_section}      }
%
%
%
%
%%%%%%%%%%%%%%%%%%%%%%%%%%%%%%%%%%%%%%%%%%%%%%%%%%%%%%%%%%%%%
%
%
Second Generation Reactive Bond Order Potential (REBO)
has been introduced by a group of authors
including D. Brenner  in \cite{brenner_REBO}. Much work has been done since 
the first  1990   paper of D. Brenner  \cite{brenner1_1990}. The main features 
of our interest here, however, remain mainly unchanged. The interaction model is
made more complex and extensive fitting is performed, so that 
the resulting REBO model interaction  is applicable to mixtures of atoms.
The structure outlined for the Tersoff model in section \ref{Tersoff_potential} is mainly applicable
also here.
%      New features added include:           
%
%
%%%%%%%%%%%%%%%%%%%%%%%%%%%%%%%%%%%%%%%%%%%%%%%%%%%%%%%%%%%%%
%
%
%
%
\subsection {        EDIP - Environment dependent interaction potential     \label{EDIP_section}      }
%
%
%
%
%%%%%%%%%%%%%%%%%%%%%%%%%%%%%%%%%%%%%%%%%%%%%%%%%%%%%%%%%%%%%
The environment dependent interaction potentials (EDIP) are first more complex 
models of interatomic interactions. The angular three body contributions and the two-body terms depend on the 
configuration of the atoms, i.e. on the number of neighbors, also known as the coordination number.
This is the characterization of the environment. The angular parts are of the 
Stillinger and Weber type. For silicon compounds
they were introduced by Bazant and coworkers in  \cite{EDIP-1} and \cite{EDIP-2}, some years later extended to carbon by Marks   \cite{Marks_carbon} and  
applied to diamond studies in  \cite{Marks_modeling_diamond} by the same author.  
With the help of switching functions depending on the coordination number
extra terms, 
dihedral rotation penalties and $\pi$-repulsion are added to the 
features of original EDIP for silicon. Due to the addition of $\pi$-orbital features, the
Stillinger Weber functionality is considerably extended, and since there is one more geometrical
feature involved, we assign to EDIP the value of 2 on the SRT-SW axis.
%
%
%%%%%%%%%%%%%%%%%%%%%%%%%%%%%%%%%%%%%%%%%%%%%%%%%%%%%%%%%%%%%
%
%
%
%
\subsection {      Long Range Carbon Bond Order Potential  - LCBOP      \label{LCBOP_section}    }
%
%
%
%
%%%%%%%%%%%%%%%%%%%%%%%%%%%%%%%%%%%%%%%%%%%%%%%%%%%%%%%%%%%%%
%
Quite recently (around 2004) a new version of interaction of Brenner bond-order type
has been introduced in references \cite{LCBOP_graphitize}, \cite{LCBOP_liquid_carbon} 
and also reported in \cite{phas_diag_carbon_LCBOP_ref}. The authors call it 
Long range Carbon Bond Order Potential, i.e. LCBOP.
This potential is thus the most recent of all the discussed ones.
It demonstrates that relatively simple model interactions can successfully model 
broad range of features when new elements of design are included. In some features
this model interaction is simpler than Tersoff-Brenner type (the angular functions), 
while it adds complexity to the coordination number dependence and the long range
part. Perhaps unfortunately, it still remains in family of interactions
where the three body angular effects (bond angles) are entered via 
pair interaction (bond order) mechanism.

%%%%%%%%%%%%%%%%%%%%%%%%%%%%%%%%%%%%%%%%%%%%%%%%%%%%%%%%%%%%%
%
%
%
%
\subsection {      General ReaxFF approach      \label{ReaxFF_section}           }
%
%
%
%
%%%%%%%%%%%%%%%%%%%%%%%%%%%%%%%%%%%%%%%%%%%%%%%%%%%%%%%%%%%%%
This interaction model was named ReaxFF, and has been
introduced as "Reactive Force Field for Hydrocarbons" in 2001   \cite{REAXFF-Hydrocarbons}.
Also ReaxFF is using the bond order concept, following Tersoff's
terminology, but the models contain much more freedom, i.e. many more parameters. 
The structure of the model is much more rich than any of the other "potential" models
discussed here. It can nearly be said that it contains elements from 
all the other models discussed.
In addition to pair interaction there are penalty functions for non-matching 
coordination number
and bond angles. The functional forms appear  mostly as additive terms, generally
they do not appear in the convoluted 
functional forms  typical for "bond order potentials" of Tersoff-Brenner type.

The ReaxFF models depend on a very large number of parameters and they provide a very realistic 
model of the chemical knowledge. The parameters are obtained in a process which is quite appropriately denoted 
by use of "training set", in analogy with the language used in non-linear optimization
of the neural networks type. The structure of the models is rather complicated, 
but it is well described in the original papers. From the original purpose as model
for hydrocarbons it has been extended to describe gradually more and more atomic combinations,
at present ReaxFF covers large portions of the periodic table.

Our classification on the RST-SW axis this method should be assigned at least number 4, since
in addition to several types of angular correlations also additional electrostatic
effects are included.

This method  should certainly be considered as an alternative to the simplest approaches above.
 On the other hand, the model is very complex (being very accurate), 
 and thus it might be too complicated for some applications which would require simplicity.
 In any case, the data collected and used in the "training" of this simulator can be very useful
 for design of simpler special purpose empirical potentials in the future.
%
%%%%%%%%%%%%%%%%%%%%%%%%%%%%%%%%%%%%%%%%%%%%%%%%%%%%%%%%%%%%%
%
%
%
%
\subsection {     Artificial neural networks  based model interactions     \label{neural_networks}    }
%
%
%
%
%%%%%%%%%%%%%%%%%%%%%%%%%%%%%%%%%%%%%%%%%%%%%%%%%%%%%%%%%%%%%
There have also been proposals to replace the model interaction
by potential functions provided by a suitable artificial neural network (ANN).
In 1999 S. Hobday  {\it et al}  \cite{neural_Brenner}     
%  Modelling Simul. Mater. Sci. Eng. 7 (1999) 397Ð412.   
%    Applications of neural networks to fitting interatomic potential functions      
%    Fit Brenner potential
investigated the feasibility of such approach by 
training a network to mimic the Brenner potential, i.e.
the energy surface was not evaluated by the Brenner formula, but
returned by the ANN, or in other words the ANN was trained on configurations where the
energy has been given by the Brenner formula. The aim of the experiment
was to test the method which in future applications would use 
not a simple and known potential, but as broad  as 
possible physical and chemical data to train the ANN.

In a recent study,  Bholoa {\it et al}
\cite{neural_TB} 
% A. Bholoa, S.D. Kenny, R. Smith, 
%Nuclear Instruments and Methods in Physics Research B 255 (2007) 1
%%%   A new approach to potential fitting using neural networks    
the ANN which gives energy surface for given local atomic positions
is trained on thousands of 
data points  for a wide range of silicon systems obtained by the tight-binding (TB) calculations.
The network had 9 nodes in the input layers, two or three hidden layers with about 
11 hidden nodes in each layer and the energy value as the output.
The authors report a very good performance of such ANN which replaces the potential form.

The idea of the method is very close to the prescription to 
use as much as possible of the accumulated knowledge to design the
model interactions. Here the authors attempted to use the known advantage
of ANNs to represent knowledge which does not have a simple logical structure,
which makes these methods suitable for character recognition, speech recognition
and many areas of nonlinear optimization.

In practical calculations this method has proved to be relatively slow when
competing with the direct Brenner formula evaluation. It would however 
probably be much faster than any ab initio method, while it could 
be trained on very thorough quantal calculations. After all, in MD context
only the resulting energy landscape is of interest, not the quantal mechanisms
themselves.
The disadvantage of this approach is similar to the ReaxFF: a very high fidelity
of the model is in principle possible, but the actual physical features of the model are in this case 
completely hidden. (In the ReaxFF there are so many features that their mutual roles 
are effectively hidden).

%
%%%%%%%%%%%%%%%%%%%%%%%%%%%%%%%%%%%%%%%%%%%%%%%%%%%%%%%%%%%%%
%
%
%
%
\subsection {     Simulations based on First Principles       \label{First_Principles_section}    }
%
%
%
%
%%%%%%%%%%%%%%%%%%%%%%%%%%%%%%%%%%%%%%%%%%%%%%%%%%%%%%%%%%%%%
The expression "first principles" is often used to classify an approach as 
the one based on only the most fundamental assumptions, and in our connection
it would often refer to inclusion of methods of quantum chemistry into 
the procedure of modeling interatomic forces and formation of the chemical
bonds (though many could object to the latter formulation, the meaning 
would be probably accepted). If taken literally, a true "first principles" approach
along these lines would be built on the following Hamiltonian which should be solved
in the framework of quantum mechanics. For a moment we keep the kinetic energies
denoted by T(r), which can be replaced by their classical as well as quantum
representations
$$
T(r_\alpha) \longrightarrow \frac{1}{2}m_\alpha\dot{r}_\alpha^2
$$
$$
T(r_\alpha) \longrightarrow -\frac{\hbar^2}{2m_\alpha}\nabla_{r_\alpha}^2
$$
schematically
$$
\sum_{nuclei\ i}  T(R_i) + \sum_{nuclei}\sum_{ i>j}\frac{Z_i Z_j}{|\vec{R}_i-\vec{R}_j|}
+ \sum_{electr\ \alpha} T(r_\alpha) 
- \sum_{nuclei\ i} \ \sum_{electr\ \alpha} \ \frac{Z_i }{|\vec{R}_i-\vec{r}_\alpha|}
$$
$$
+  \sum_{electr }  \sum_{\alpha>\beta} \frac{1}{|\vec{r}_\beta-\vec{r}_\alpha|}
$$
%sum over all nuclei T(R)
%sum over all nuclei sum over all nuclei Zi Zj /(Ri -  Rj)
%sum over all nuclei ( sum over all electrons ) - Zi / ( Ri - r alpha )
%( sum over all electrons ) ( sum over all electrons )  + 1 / ( r alpha   - r beta )
             
The summations over electrons can be grouped into the atoms and
Born-Oppenheimer type approximations can then be used much like in the so called 
semiclassical collision theory for atom-atom or molecular collisions. Further,  a series of 
well defined and controlled approximations could be carried out further to 
arrive at different simulation methods (a useful review has been given by Marx and Hutter \cite{ab-initio}).

Following Kohn's Nobel lecture \cite{kohn_nobel} , one can ask if such ambitions are in fact
fruitful. We know that chemistry does not violate any physical first principles, but
we also know that many situations which chemistry describes may become extremely dependent
on from the point of physics more or less surprising accidental dependences on
geometry, accidental quantal energy degeneracies and many other effects which are not possible
to be seen from the formula above. Thus, one should always have in mind that quantum chemistry
can never fully replace the laboratory and the ever richer toolboxes of physical chemistry.

The purpose of this statement is to realize that a really fruitful future
of the empirical potentials, including an obvious advantage over the 
'more fundamental' methods is to build empirical interactions which are based
on all accessible relevant chemical and physical data instead of on
requirements of simple functional form or derivability 
by complex approximations from a 'more fundamental' formulation.

In quantum chemistry, the above Hamiltonian with various approximations is 
solved using various types of selfconsistent field approaches, recently mostly based on
density functional theory (DFT), earlier on Hartree-Fock (HF) approaches. 
Historically, a very succesful approaches have been based on 
linear combinations of atomic orbitals, LCAO. 
In solid state physics the LCAO approaches have been adapted to the so called 
tight binding approximation (TB), where the geometry could enter in a very simplified
form, which was very useful both for qualitative understanding as well as 
numerical evaluation of the electron energy band structure (e.g. the well known 
work of Slater and Koster \cite{TBMD_LCAO_Slater_Koster}).

In fact, the "first principles" methods can be built on much less fundamental principles,
approximative methods like the tight binding method or density functional theory are used.
In the literature both the DFT methods and the TB methods are referred to as 
{\it ab initio} methods.
%
%%%%%%%    For our review here, the resulting 
%
%%%%%%%%%%%%%%%%%%%%%%%%%%%%%%%%%%%%%%%%%%%%%%%%%%%%%%%%%%%%%
%
%
%
%
\subsection {        ABOP - Analytic Bond Order Potentials     \label{BOP_section}      }
%
%
%
%
%%%%%%%%%%%%%%%%%%%%%%%%%%%%%%%%%%%%%%%%%%%%%%%%%%%%%%%%%%%%%
%
%
The  Analytic Bond Order Potentials  (ABOP) are reviewed in ref. \cite {ABOP_review}.
This approach follows the work of Pettifor and Oleinik \cite{ABOP_Pettifor_Oleinik},
who attempt to go beyond the Tersoff's assumptions,  and it can be described 
to be in fact based on the tight binding approximation,  
\cite{TBMD_LCAO_Slater_Koster} \cite{TBMD_Tomanek} \cite{TBMD_Cleri}.
The tight binding molecular dynamics is a large field with lots of locally adopted concepts 
which might sometimes make it difficult to follow for workers outside of the field.
The geometrical features are included in a really fundamental manner, building on the
angular dependence of the exchange or hopping matrix elements. However, the
formalism of the derived model interactions is rather complex and will not be reviewed here. 
The review in ref. \cite {ABOP_review} can be consulted for details.
 
%    {\it Descibe the matrix elements. Describe the LCAO aspect of TBMD. Give references to 
%    Solid state physics}. 
The ABOP thus attempts to include consistently an approximation to the quantum chemistry
and on our RST-SW axis it should be given at least index 3, since in principle all the 
aspects of the geometry following from quantum theory are included. 
They should certainly be considered as an alternative to the simplest approaches above, but 
on the other hand they might be too complicated for some applications.

%
%
%
%
%
%
%%%%%%%%%%%%%%%%%%%%%%%%%%%%%%%%%%%%%%%%%%%%%%%%%%%%%%%%%%%%%
%
%
%
%
\section {       Tersoff and Brenner Potentials        \label{tersoffbrenner_papers}    }
%
%
%
%
%%%%%%%%%%%%%%%%%%%%%%%%%%%%%%%%%%%%%%%%%%%%%%%%%%%%%%%%%%%%%
As discussed above, these potentials were originally suggested and explored by Tersoff
and later modified and extended  by D. Brenner and coworkers with a wide selection
of applications. To some small degree the functional forms but mainly the parameters were 
modified to suit many different systems.
There are hundreds of works using one or other form of these potentials, as well as many
further modifications and adjustments. A full review of all these attempts is virtually
impossible to be carried out in a reasonable format, however it might 
be illustrative to select some examples.
One of the examples can be the use of Brenner potential for simulations of 
the self-assembly of fullerenes (1998), carried out by 
 Yamaguchi and  Maruyama\cite{Maruyama_FULLEREN}
%    Y. Yamaguchi and S. Maruyama, Chem. Phys. Lett. 286 (1998) 336
%      A molecular dynamics simulation of the fullerene  formation process
%     ({\it Japanese reference - and review to be added} ).

We will however, try to analyze these potentials, which gave the name to the whole
method, as "bond-order potentials". As seen from the above section, not all of the
empirical potentials are necessarily connected with this particular interpretation of the
bond order concept.
%%%%%%%%%%%%%%%%%%%%%%%%%%%%%%%%%%%%%%%%%%%%%%%%%%%%%%%%%%%%%
%
%
%
%
\subsection {      Many body character of Tersoff interaction      \label{tersoff_manybody}           }
%
%
%
%
%%%%%%%%%%%%%%%%%%%%%%%%%%%%%%%%%%%%%%%%%%%%%%%%%%%%%%%%%%%%%
In this section 
we attempt rewriting the Tersoff potential in a general form, as 
a many-body potential. Why are they not additive as SW or ReaxFF? And why is
the functional form always an exponentials?
%
%
%%%%%%%%%%%%%%%%%%%%%%%%%%%%%%%%%%%%%%%%%%%%%%%%%%%%%%%%%%%%%
%
%
%
%
The many-body interaction in Tersoff approach is in fact written in the
following general form
\[ V ( r_1, r_2, \ldots \ldots r_N ) = \frac{1}{2} \sum^N_i \sum^N_{j ; j \neq
   i} F_{ij} \left( \sum^N_{k \neq i, j} G ( r_i, r_j, r_k ) \right) 
\]
One can imagine that this form is a result of a certain summation of this type
of series:
\begin{eqnarray}
   V ( r_1, r_2, \ldots \ldots r_N ) 
   & = &
           \sum^{( pairs )}_{i, j ; i < j}
   P_{ij} \left( r_i, r_j \right) + \sum^{( triplets )}_{i, j,
   k} T_{ijk} ( r_i, r_j, r_k )  \nonumber \\
   & \ \ & 
          + 
          \sum^{( quadruplets )}_{i,j, k, m} Q_{ijkm} ( r_i, r_j, r_k, r_m ) + \ldots
\end{eqnarray}   
in the sense described by Stillinger and Weber. Only special types of the
latter general expansion when summed would result into the former type of
expression. It is thus a model assumption, based on the idea of 'bond
order', i.e. the variable strength of the interaction, which in Pauling's
empirical formula \cite{pauling}
% L. Pauling. J. Am. Chem. Soc. 69 (1947), p. 542.
is related to the length of the bond. 
%G. C. Abell, Phys. Rev. B 31 (1985) 6184    %   - 6196 
% Empirical chemical pseudopotential theory of molecular and metallic bonding
In his work on empirical chemical pseudopotentials for metallic bonding 
Abell 
\cite{abell} further generalized (somewhat arbitrarily) this dependence to  
reflect also  the environment of the two bonded atoms.

In the ABOP formulation, the mentioned shape is attempted to be derived from a
summation, using a certain type of Green's function formulation. The ABOP potentials
are based on TBMD applications, modeling the diagonalization of
sparse matrices. This is a very appealing approach, since it provides
to some degree a sound theoretical basis for the whole bond-order approach,
as discussed in the section \ref{BOP_section}
%
%%%%%%%%%%%%%%%%%%%%%%%%%%%%%%%%%%%%%%%%%%%%%%%%%%%%%%%%%%%%%

\subsection {The geometry treatment in the Tersoff Potentials \label{tersoff_geometry}}

The geometry treatment is in fact of the same type as that of Stillinger - Weber, i.e. simply the preferred angle
is contained in the  function $g(\theta_{ijk})$,
where $\theta_{ijk}$ is the angle between the lines connecting the three atoms 
%
%
%%%%%%%%%%%%%%%%%%%%%%%%%%%%%%%%%%%%%%%%%%%%%%%%%%%%%%%%%%%%%%%%%%%
$$ g(\theta) = 1 + \left(      \frac{c}{d}\right       )^2 
                 - 
                  \frac{c^2}{d^2 + [ h - \cos  \theta  ]^2}  
\ \ \ \ \ \ \ \ \ \ \  \ \ \  
$$
%%%%%%%%%%%%%%%%%%%%%%%%%%%%%%%%%%%%%%%%%%%%%%%%%%%%%%%%%%%%%%%%%%%
%
%
The value of $h=-0.598$ given in Tersoff's paper selects the bond angle
$126.7^o$, while in Brenner's case this is sometimes replaced by $180^o$
(this is probably only an omission in the paper, listing only one value).
%
%%%%%%%%%%%%%%%%%%%%%%%%%%%%%%%%%%%%%%%%%%%%%%%%%%%%%%%%%%%%%
%
%                         r-----------------n
%                         I  / \    / \     I    FIGURE  
%                         I  | |    | |     I
%                         ^-----------------^
%
\begin{figure}[htb]
\center{
\includegraphics[height=3.5cm]{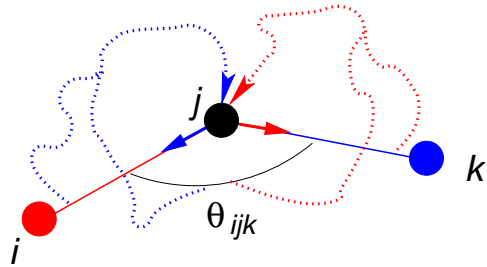}
}
 \caption{This figure attempts to illustrate the Tersoff basic method. The interaction
 appears as a two body potential, but the strength parameters for the pair i-th and j-th atom
 are calculated using the angle of (all) k-th atom(s). The dependence on the angle between the
 bonding lines enters in a complicated way into the evaluation of the attractive part
 of the interaction between i-th and j-th atom (cf. eq. \ref{tersoff-angle})
    }
 %%%
\label{bond-angle-3}
\end{figure}
%
%
%
%%%%%%%%%%%%%%%%%%%%%%%%%%%%%%%%%%%%%%%%%%%%%%%%%%%%%%%%%%%%%
Clearly, only one angle can be included. Which then poses a question - how can Tersoff-Brenner 
potentials model both graphenes and diamonds.
%%%%%%%%%%%%%%%%%%%%%%%%%%%%%%%%%%%%%%%%%%%%%%%%%%%%%%%%%%%%%
%
%                         r-----------------n
%                         I  / \    / \     I    FIGURE  
%                         I  | |    | |     I
%                         ^-----------------^
%
\begin{figure}[htb]
\center{
\includegraphics[height=5.5cm]{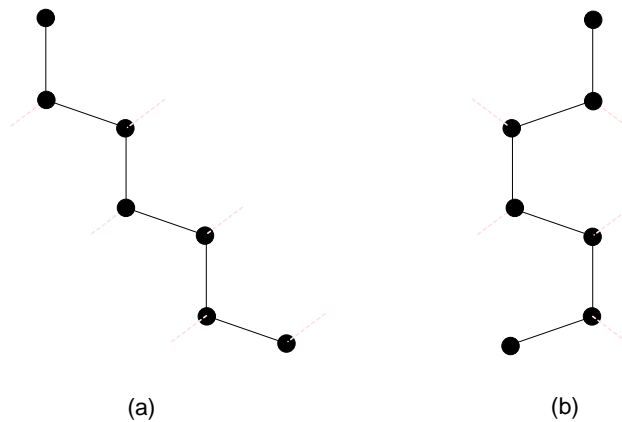}
}
 \caption{Diamond (a) and hexagonal lonsdaleite (b)    }
 %%%
\label{diamond-vs-lonsdale-plane1}
\end{figure}
%
%
%
%%%%%%%%%%%%%%%%%%%%%%%%%%%%%%%%%%%%%%%%%%%%%%%%%%%%%%%%%%%%%  four_particles
The answer is that it is not the potential itself, but  the
interplay of the potential and the properties of the space.
It is possible to change a little bit the parameters,
but the same potential only allows one angle.

% Different story is the 4-body arrangement.
% Here the distances (physically correctly) prefer the diamond to lonsdaleite.

The bond order strategy can be characterized as follows: the effect of the angular
correlation is expressed by influencing the bond strength (order) of the bonds formed by other
pairs. The three body correlation characterized naturally by the angle between
the bonds is  in BOP instead taken into account by expressing the strength of 
both neighbouring  bonds as rather complex functions of the angle between the bonds.
Adjusting these functions and their parameters to chemical data and results of 
quantum chemical calculations is thus quite complicated procedure.
This is in contrast to the simple form of angular dependence in the Stillinger and Weber
which is expressed as an additive term.

%
%%%%%%%%%%%%%%%%%%%%%%%%%%%%%%%%%%%%%%%%%%%%%%%%%%%%%%%%%%%%%%%%%%%
%
%
%
%
\subsection {Tersoff-Brenner potential and four particle geometry\label{effective-correlation}}
%
%
%
%
%%%%%%%%%%%%%%%%%%%%%%%%%%%%%%%%%%%%%%%%%%%%%%%%%%%%%%%%%%%%%%%%%%%
%
A given carbon atom both in the Diamond and Lonsdaleite structures has up the second
neighbors completely identical
neighborhoods, the differences appear first by the third neighbors as can be understood from  
figure \ref{diamond-vs-lonsdale-plane1}.
The interaction with the nearest neighbors
can be sufficiently well represented by a three-body
interaction of as simple type as the Stillinger-Weber type.
(Tersoff-Brenner potentials contain in principle more complex correlations). 
To differentiate between the two different 4-atom conformations
(aliphatic type and "boat chains" - see again the figure \ref{diamond-vs-lonsdale-plane1}) in the
two discussed structures,
a 4-body correlation must be made effective in the interaction model.

Due to the implicit sum over all triplets in the bond strength
of the Tersoff-Brenner potentials, one would
expect the possibility to have a 4-body correlation.
However, with the cut-off function used in the standard
formulation, the mutual influence is limited to the nearest
neighbors, the standard cut-off prevents any higher than 3-body
correlation. This is not inherent limitation of the model, it is simply the choice of the 
cut-off parameter.

If the repulsion part (cf. equation \ref{eq84}) would be allowed to 
act over a longer range (using a different cut
off for each of the two terms), one could in principle model
a four body correlation (using only 3-body interactions)
without any other modifications of the Tersoff-Brenner model.  
One should realize that different cut-off could lead to
a little positve energy (repulsive) region, appearing as a little barrier
in the two atom case.

Without such type of modification, both Tersoff-Brenner and Stillinger-Weber
interactions lead to a situation where both diamond and lonsdaleite are
energetically completely equivalent.
%%%%%%%%%%%%%%%%%%%%%%%%%%%%%%%%%%%%%%%%%%%%%%%%%%%%%%%%%%%%%%%%%%%
%
%
%
%
    \subsection {Investigation methods \label{explore_angle}}
%
%
%
%
%%%%%%%%%%%%%%%%%%%%%%%%%%%%%%%%%%%%%%%%%%%%%%%%%%%%%%%%%%%%%%%%%%%   Mathematica \cite{mathematica} Maple \cite{maple}
%     thesis \cite{suhail2003}    MATLAB \cite{matlab} 
We have developed several simple techniques to study the angular aspects. 
For this purpose we have written small tools for several mathematical systems.
In order to be able to perform quickly evaluation of forces
from potentials, we have developed a code in Mathematica \cite{mathematica} (to allow even the 
inspection of the analytic form of forces), as well as in Maple \cite{maple}. However,
the advantages of the analytic form are shadowed by the complexity
and length of the expressions, so the usefulness of this approach is limited
in this context (some examples can be found in the thesis \cite{suhail2003}). 
More recently we have used MATLAB \cite{matlab}, which is much more suitable for numerical inspection.
In this connection we have used the 
ability to perform mathematical 
operations on complicated structured objects 
in one single statement using ordinary mathematical notation. 
Thus an evaluation of potentials or forces which would in other computer languages require a whole
special computer program can be not only evaluated, but also
visualized in a couple of lines which are directly interpreted. 
The MATLAB work is available  as preprint \cite{CPC_paper} and submitted for publication.
(Open source systems GNU-Octave and SCILAB can have very close
syntax and functionality so that some of our shorter scripts for MATLAB can also be 
applied using these two free systems.)
%   What have we learned: ......

%%%%%%%%%%%%%%%%%%%%%%%%%%%%%%%%%%%%%%%%%%%%%%%%%%%%%%%%%%%%%%%%%%%
%
%
%
%
%  \section {Extensions of Tersoff and Stillinger-Weber models \label{fourbody}}
%
%
%
%
The functional dependence of the empirical potentials discussed is not really 
given by any very thorough tests. A convenient form has been chosen at a certain
point and the information is carried by the parameters. The question about how optimal
the functional shapes really are has not been raised. %%%%%%%%%%%%%%%%%%%%%%%%%%%%%%%%%%%%%%%%%%%%%%%%%%%%%%%%%%%%%
\subsection {      Form and cut-off of Tersoff potential     \label{tersoff_cutoff}           }
\begin{figure}[htp]
%\centering
\begin{tabular}{cc}
\begin{minipage}{0.450\linewidth}
%\centering
\includegraphics[scale=0.42]{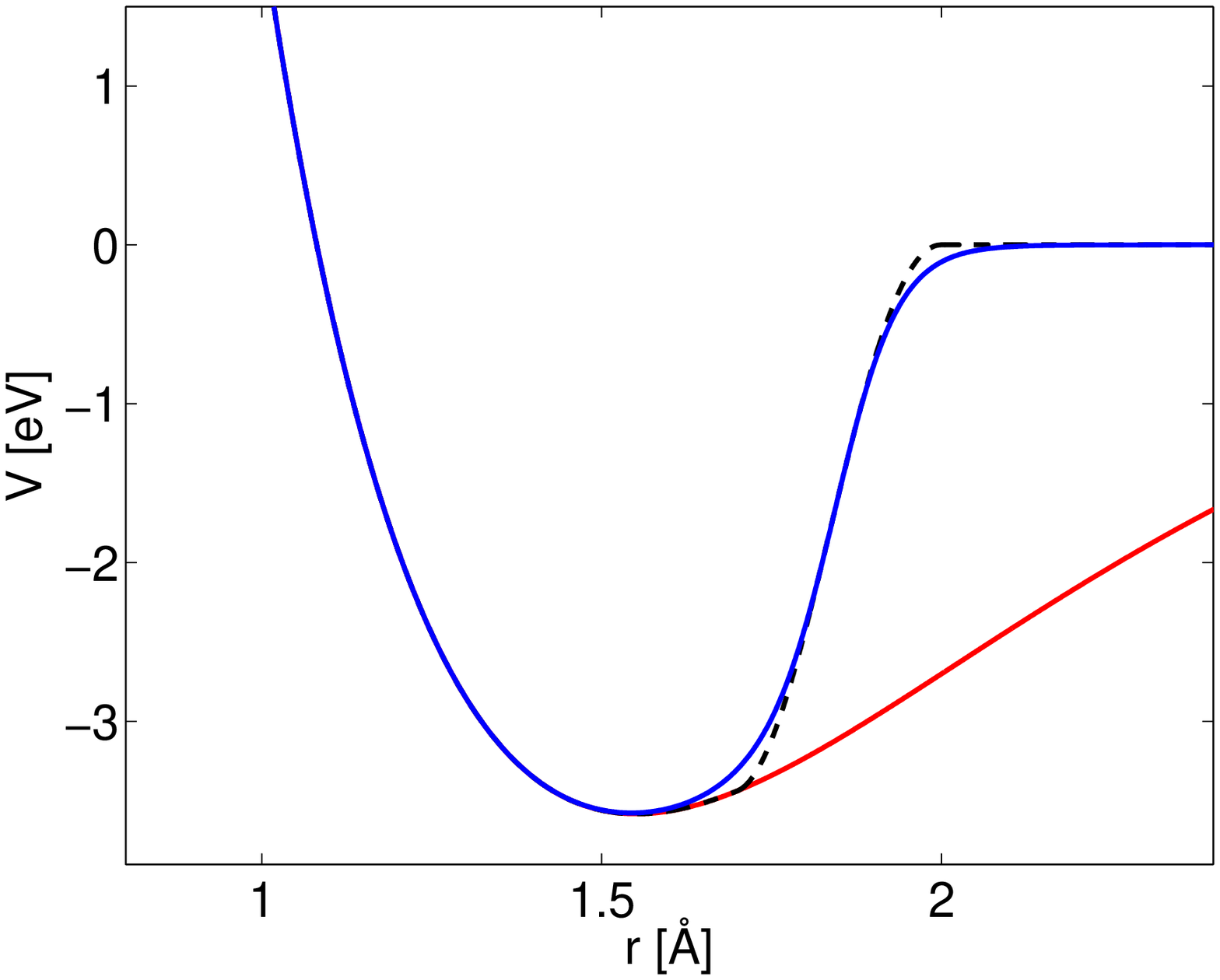}
\end{minipage}
&
\begin{minipage}{0.450\linewidth}
%\centering
\includegraphics[scale=0.42]{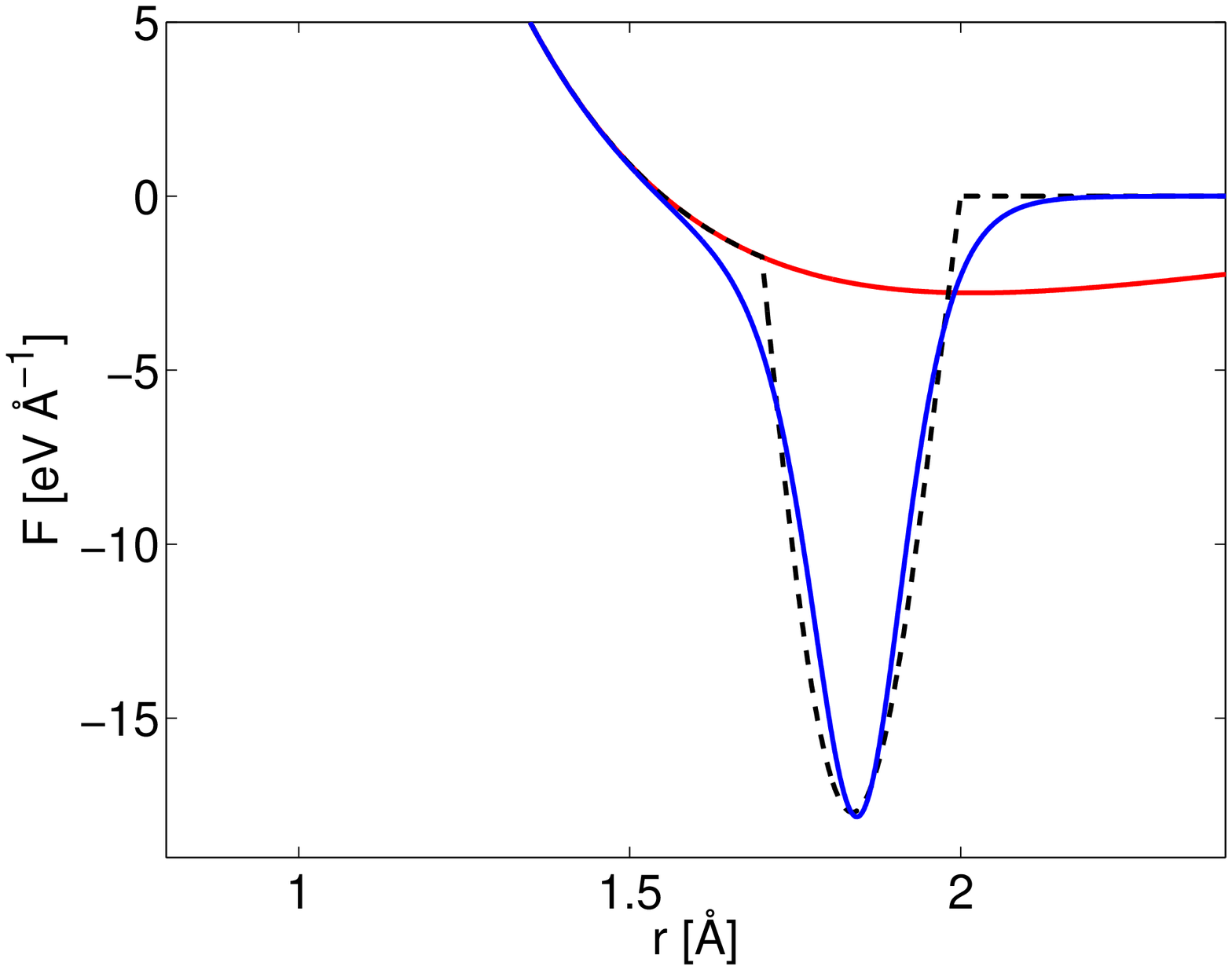}
\end{minipage} \\
\begin{minipage}{0.450\linewidth}
\centering
(a) 
\end{minipage}
&
\begin{minipage}{0.450\linewidth}
\centering
(b) 
\end{minipage}
\end{tabular}
\caption{The potential and force with Tersoff-Brenner and Fermi cut-off functions;
(a) Potential shape without any cut-off in solid red, with Tersoff-Brenner cut-off in dashed black, 
with Fermi cut-off in solid blue; 
(b) The corresponding forces with the same notation as in (a)}
\label{fermi-basis}       %\label{Potentials3Forces3}
\end{figure}

\begin{figure}[htp]
%\centering
\begin{tabular}{cc}
\begin{minipage}{0.450\linewidth}
%\centering
\includegraphics[scale=0.42]{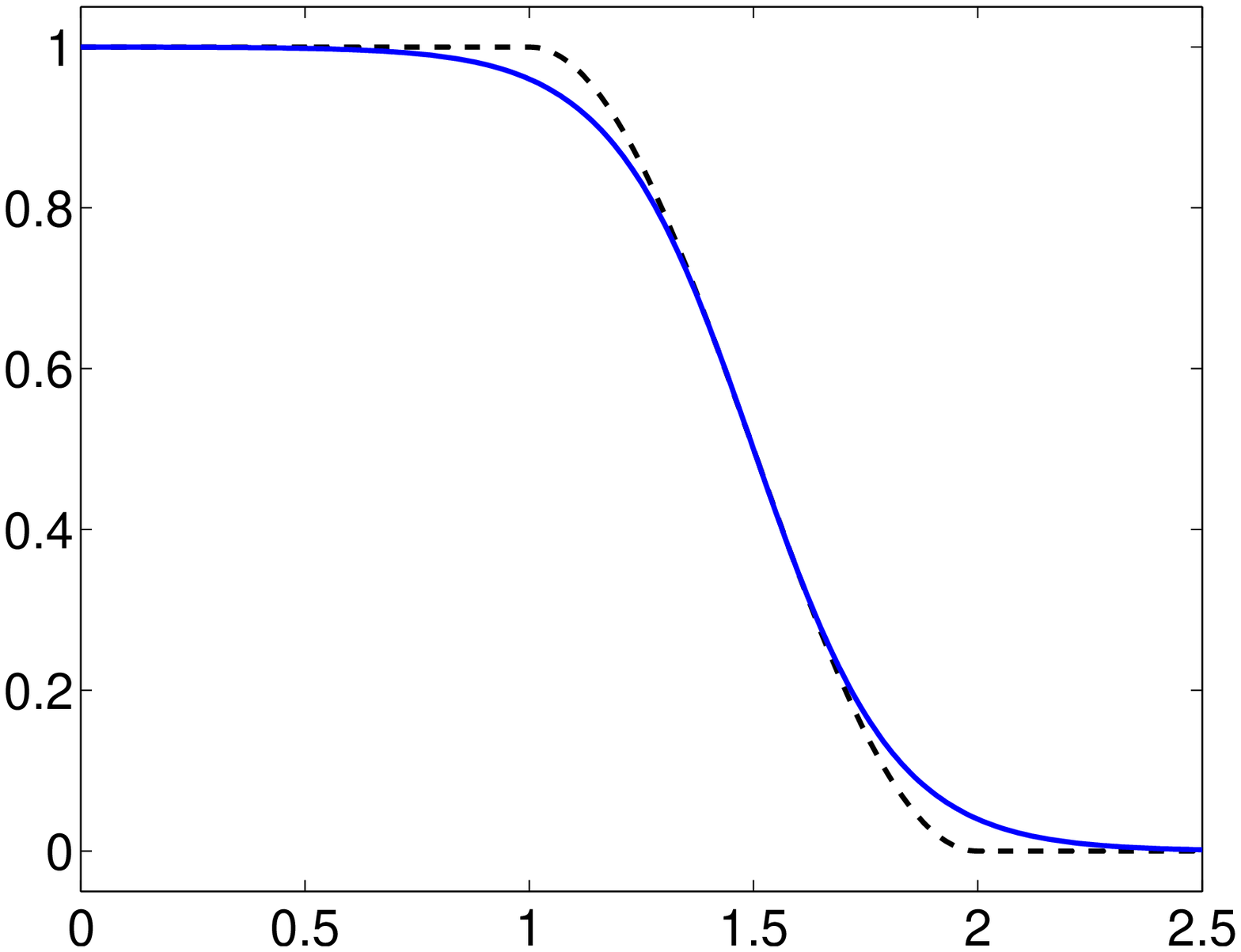}
\end{minipage}
&
\begin{minipage}{0.450\linewidth}
%\centering
\includegraphics[scale=0.42]{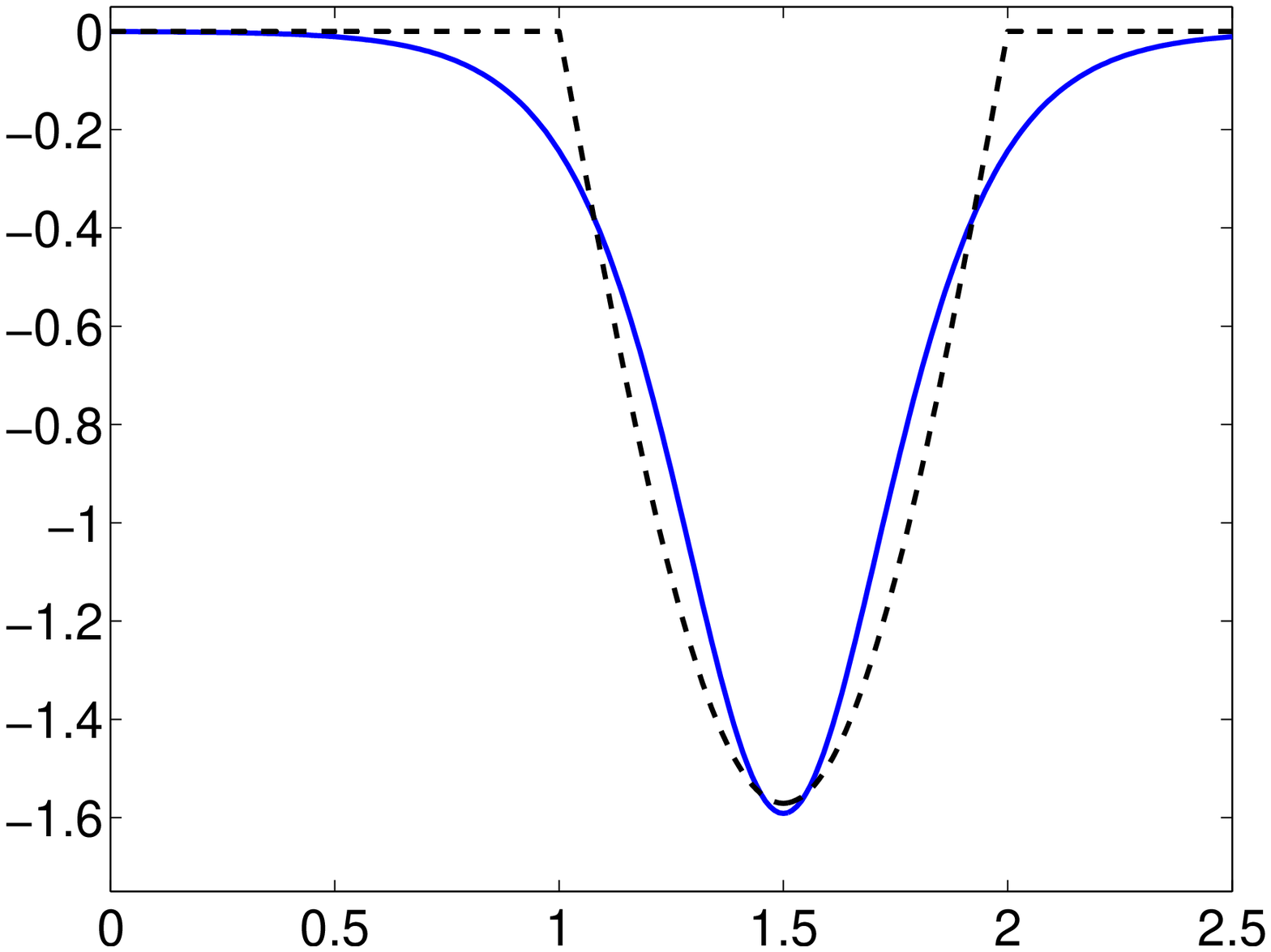}
\end{minipage} \\
\begin{minipage}{0.450\linewidth}
\centering
(a) 
\end{minipage}
&
\begin{minipage}{0.450\linewidth}
\centering
(b)
\end{minipage} 
\end{tabular}
\caption{Comparison between Tersoff-Brenner and Fermi cut-off functions;\newline
(a) Comparison of the cut-off functions; Tersoff-Brenner in dashed black, Fermi in solid blue
; (b) Comparison of the derivatives of cut-off functions: notation as in (a)}
\label{fermi-tersoff}   %    \label{Potentials3Forces3}
\end{figure}

%
%
%
%
%%%%%%%%%%%%%%%%%%%%%%%%%%%%%%%%%%%%%%%%%%%%%%%%%%%%%%%%%%%%%
For Tersoff-Brenner potentials, most of the
active forces are in fact provided by the cut-off function, as the figure \ref{fermi-basis} shows.
Stillinger-Weber use a different, simpler function, which could easily be used also by 
Tersoff-class of potentials.
%
%%%%%%%%%%%%   EXTRA1  %%%%%%%%%%%%
%
%%
%%
%
%%%%%%%%%%%%%%%%%%%%%%%%%%%%%%%%%%%%%%%%%%%%%%%%%%%%%%%%%%%%%  four_particles.pdf
The cut-off function used by all Tersoff-Brenner applications
is the cosine-type cut-off formula eq. \ref{eq85}  which does not have a smooth derivative.
In all our work we have replaced it by 
the Fermi function $( \exp((r-r_0)/d) + 1)^{-1}$ which is 
smooth everywhere and the parameter $d$ can be easily adjusted. Figure \ref{fermi-tersoff}
shows the comparison of the two cut-off treatments.
%
%%%%%%%%%%%%   EXTRA2  %%%%%%%%%%%%
%%
%%

%
%%%%%%%%%%%%%%%%%%%%%%%%%%%%%%%%%%%%%%%%%%%%%%%%%%%%%%%%%%%%%  four_particles.pdf

%%%%%%%%%%%%%%%%%%%%%%%%%%%%%%%%%%%%%%%%%%%%%%%%%%%%%%%%%%%%%%%%%%%
%
%
%
%
\subsection {Extending Stillinger-Weber approach by four-body correlations. \label{fourbody_SW}}
%
%
%
%

%%%%%%%%%%%%%%%%%%%%%%%%%%%%%%%%%%%%%%%%%%%%%%%%%%%%%%%%%%%%%

\begin{figure}[htb]
\center{
\includegraphics[height=7.5cm]{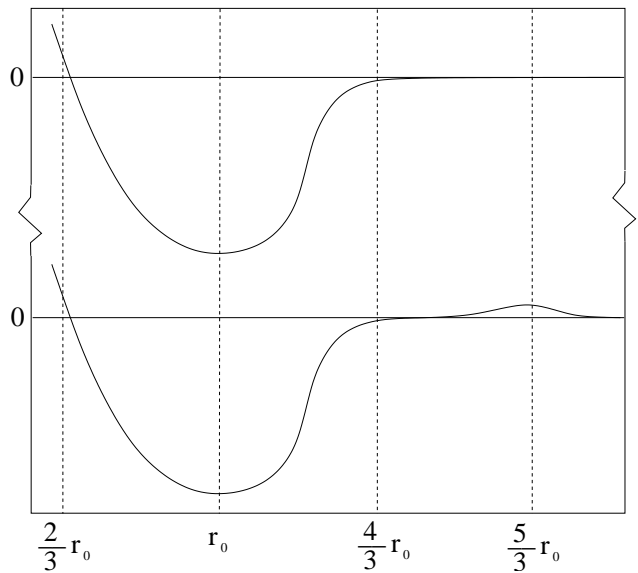}
}
 \caption{ Schematic representation of the two-body part of S-W potential
 accounting for the four-body correlations. Upper part: the usual shape of 
 Tersoff and SW potentials; Lower part: in the region of lonsdaleite third neighbor
 distance there appears a little repulsion energy. 
  }
 %%%
\label{bumpy-2-body}
\end{figure}

The four-body correlation energy, usually referred to as dihedral angle potential, 
could be added to the Stillinger-Weber model by introducing a  four body term, 
formally completely analogous to the allready present three-body term.
However, as discussed in the section 
\ref{effective-correlation}  above, one could also 
attempt to add an effective four-body correlation mechanism by only choosing suitable
modifications of the existing model.
%%%%%%%%%%%%%%%%%%%%%%%%%%%%%%%%%%%%%%%%%%%%%%%%%%%%%%%%%%%%%%%%%%%
%
%
%
%
We want to concentrate on the question which can be formulated as that
of difference between diamond and lonsdaleite, or in much more complicated way as the
dihedral angle dependence.
With only three-body correlations taken into account, Stillinger-Weber potentials 
will lead with more  or less equal probability to both cubic and hexagonal
lattice formation. On the other hand, a realistic potential should be able to distinguish between 
these arrangements in plane
shown in fig. \ref{diamond-vs-lonsdale-plane1}. As discussed earlier in Sec. \ref{effective-correlation}, 
Tersoff-type potentials could do that in principle with 
a modified cut-off treatment.
The same effective four-body effect
due to a modified two-body interaction can also be extended
to the Stillinger-Weber model without introducing a new
explicit 4-body term. This can be done by modifying the functions defining the potential shapes
defined by equations \ref{stillinger_f_2} and
\ref{stillinger_h_func}.
The schematic representation of this type of potential is given in
figure \ref{bumpy-2-body}. It shows the usual shape of 
 Tersoff and SW potentials, but also with added a small repulsion region
close to the lonsdaleite third neighbor distance. The physical origin of this
 model term is electron-electron repulsion not accounted for by the 
 independent electron picture implicit in the discussed models. 
%
%
%%%%%%%%%%%%%%%%%%%%%%%%%%%%%%%%%%%%%%%%%%%%%%%%%%%%%%%%%%%
%
%
%
%
%
%%%%%%%%%%%%%%%%%%%%%%%%%%%%%%%%%%%%%%%%%%%%%%%%%%%%%%%%%%%%%  four_particles.pdf
With so modified basic two-body interactions lonsdaleite 
becomes energetically less favorable than diamond, which
is a desirable result, both from observation and from calculations
\cite{First-principlesDiamonds}, \cite{HexagonalPolytypes}.

It is interesting to note that a similar situation is encountered in the hydrocarbons, the alkanes.
The diamond-like arrangement is also there energetically preferred. The isomers which have the 
lonsdaleite analogue structure have a higher energy denoted as so called bond strain. 
In this case the decreased binding is ascribed to the electron-electron repulsion between electron
bonding pairs present on any two hydrogens in the boat configuration, fig.
\ref{diamond-vs-lonsdale-plane1}b. 
%
%%%%%%%%%%%%%%%%%%%%%%%%%%%%%%%%%%%%%%%%%%%%%%%%%%%%%%%%%%%%%
%
%
%
%
\section {      Conclusion    \label{conclusion}           }
%
%
%
%
%%%%%%%%%%%%%%%%%%%%%%%%%%%%%%%%%%%%%%%%%%%%%%%%%%%%%%%%%%%%%
We have discussed the various approaches 
to reactive model interactions found in literature. We have illustrated 
how varied these approaches in fact are in several important aspects.
We have classified qualitatively the interactions by a position
on the RST-SW axis introduced in section \ref{stillinger_weber_sect}, according to the attempted simulation of the geometrical features.

%    and illustrated
%    the question if the bond order concept really 
%    is best suited for the design of the models of interaction. The answer we leave to the reader,
%     illstration of the various aspects seemed to be more useful then giving marks.
%    The 1993 paper on "Born-Oppenheimer Molecular dynamics" seems to define a better perspective.
%    The PES should be modeled, without the in many respects not precisely defined 
%    "bond order", a concept which is alternating between "bond strength", "potential" 
%    and a combination of electron densities.

The recent LCBOP discussed in subsection \ref{LCBOP_section} shows that new 
ideas may still contribute to a usability of relatively simple models.
Though the computing progress makes the so called {\it ab initio} methods
increasingly more feasible, simple model interactions might still be very useful for certain types
of studies. This is also supported by our discussion of possible
inclusion of four-body or dihedral angle effects in the basically three-body models.

Hybrid methods, combining e.g. DFT parts which could provide re-defined parameters
in well designed simple model interactions of generalized EDIP-type
 might become elements  of increasing importance as new groups of researchers 
will use MD-based approaches to address possibly new questions about 
interest in material sciences and nanotechnological applications.

%    We conclude that even in spite of computational development the empirical potentials
%    will remain useful for studies of some aspects of atomic structures also in the
%    future. 

We conclude that there are large variations in the methods to design the empirical interactions,
it does not seem possible to assume that all possible simple approaches have been considered,
since the mentioned recent multiple minima isotropic pair potential method remained undiscovered until
quite recently. Though this particular method does not have a direct practical importance,
we find its recent discovery as an indication that possible simpler alternatives to 
the existing models should be explored. 
 This paper also forms a basis for our work with new
simple interaction models \cite{OBMD_paper} which is presently submitted for publication.
%%%%%%%%%%%%%%%%%%%%%%%%%%%%%%%%%%%%%%%%%%%%%%%%%%%%%%%%%%%%%%%%%%%
%
%
%
%
\section*{Acknowledgments}
We would like to thank Dr Norbert L{\"u}mmen at University of Bergen for
very helpful and enlightening discussions of his work based on the ReaxFF system.
%
%
%
%

%%%%%%%%%%%%%%%%%%%%%%%%%%%%%%%%%%%%%%%%%%%%%%%%%%%%%%%%%%%%%

%%%%%%%%%%%%%%%%%%%%%%%%%%%%%%%%%%%%%%%%%%%%%%%%%%%%%%%%%%%%%%%%%%%%%%%%%%%%%%%%%%%%
\section*{References}
%%%%%%%%%%%%%%%%%%%%%%%%%%%%%%%%%%%%%%%%%%%%%%%%%%%%%%%%%%%%%%%%%%%%%%%%%%%%%%%%%%%%%

\end{document}